\documentclass{moriond} 
\usepackage{wrapfig}
\usepackage{lipsum} 
\usepackage{amsmath}
\usepackage{comment}

\def\be{\begin{equation}}
\def\ee{\end{equation}}
\def\bea{\begin{eqnarray}}
\def\eea{\end{eqnarray}}

\newcommand{\Photo}{\includegraphics[height=45mm]{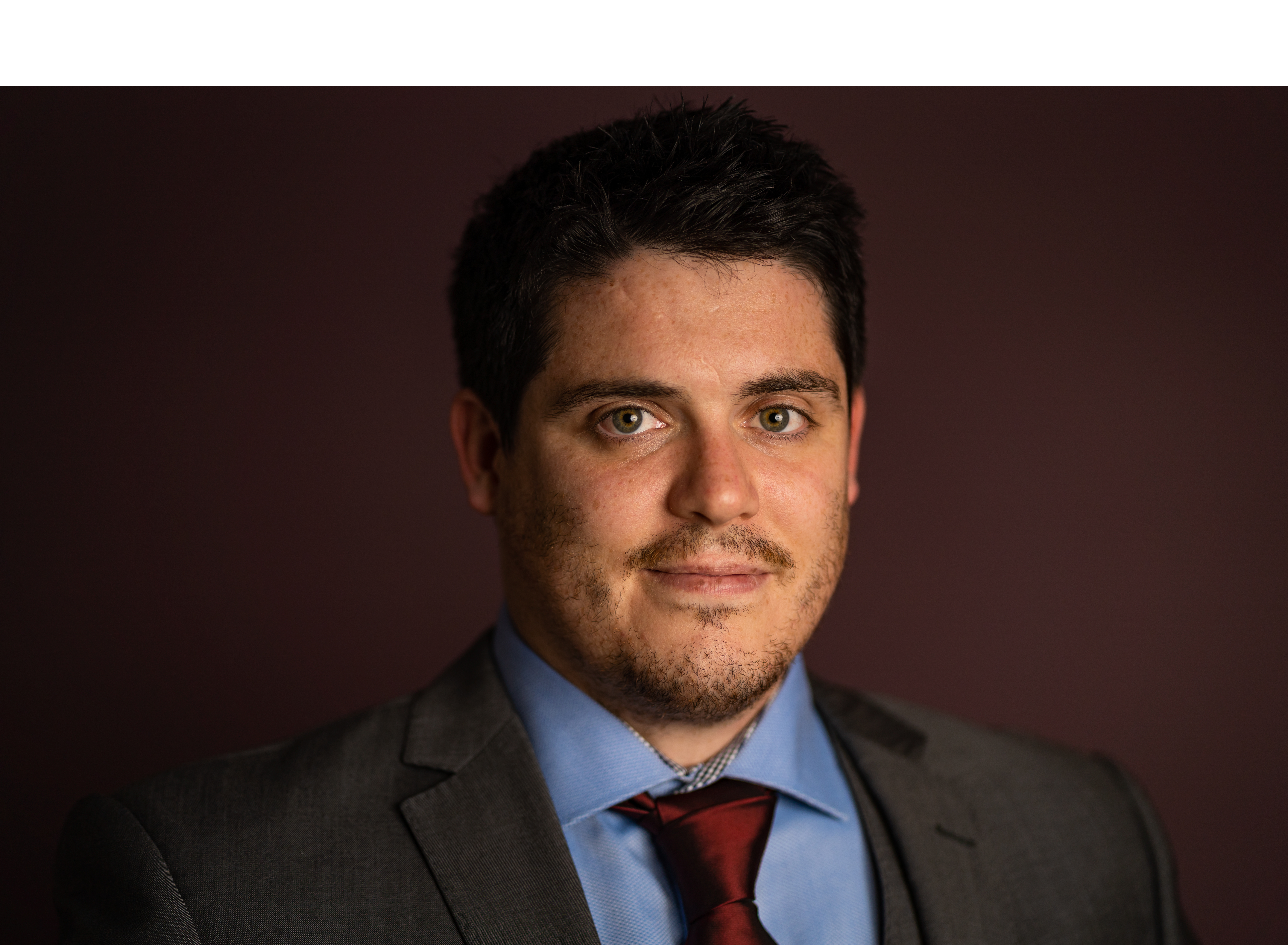}}
\begin{document}

\vspace*{4cm}

\title{Optimisation of table-top 3D interferometers for Observational Quantum Gravity}

\author{W. L. Griffiths, L. Aiello, A. Ejlli, A. L. James, S. M. Vermeulen, K. L. Dooley, and H. Grote }

\address{Gravity Exploration Institute, School of Physics and Astronomy, \\Cardiff University, CF24 3AA, UK}

\maketitle
\abstracts{With the use of twin, co-located, 3D interferometers, Cardiff University's Gravity Exploration Institute aims to observe quantum fluctuations of space-time as predicted by some theories of quantum gravity. Our design displacement sensitivity exceeds that of previous similar experiments, which have constrained the magnitudes of the fluctuations in the 1--25 MHz band. The increased sensitivity comes in large part from the comparably higher circulating power we aim to achieve, which reduces the overall shot noise. One complication of higher circulating power is an increase in contrast defect light, which includes higher-order modes. We will use the DC-readout scheme, whose dark-fringe offset must sufficiently dominate the contrast defect in order to detect faint signals. However, too much total output power risks saturating the high-bandwidth photodetectors. Suppressing the higher-order mode content of the contrast defect is a key strategy to realising the high circulating power and eliminating non-signal-carrying power that contributes to shot noise. For this, the inclusion of an output mode cleaner, whose design is described, is required.}

\section{Introduction}\label{sec:Intro}
An expected consequence of theories of quantised space-time is an irreducible variance of the fluctuations of repeated measurements of distance~\cite{hogan_measurement_2008,amelino-camelia_quantum_2013}. The holographic principle~\cite{bousso_holographic_2002} implies that these distance fluctuations are correlated in a given volume of space-time such that the cross-spectrum of the respective differential length changes in two identical and physically overlapped (co-located) interferometers, would reveal any coherent signals. There is also theoretical evidence that the quantum space-time fluctuations exhibit certain angular correlations~\cite{vermeulen_et_al_experiment_2021}, hence the 3D configuration. Signatures of quantum space-time fluctuations will thus present themselves as a constant, unavoidable and common noise between the two instruments. 

Co-located interferometery experiments for quantum gravity investigations are not new: the Fermilab Holometer~\cite{chou_et_al_holometer_2017} constrained the magnitude of longitudinal quantum space-time fluctuations~\cite{chou_et_al_interferometric_2017}. Each of their interferometers had a shot-noise-limited displacement sensitivity of $10^{-18}$~m~/~$\sqrt{\text{Hz}}$ in the 1--25~MHz band. With a five-fold increase of circulating power (among other improvements, including the injection of squeezed light), we expect to exceed this displacement sensitivity and increase the detection bandwidth up to about 250~MHz.  

These proceedings motivates and describes the technical design of our output mode cleaner, a new component that is essential for achieving the higher circulating power. For a thorough description of the overall experimental aims, design and projected sensitivity, we recommend our recent paper published in CQG~\cite{vermeulen_et_al_experiment_2021}.

\section{Experimental design}\label{sec:exp design}

\begin{figure}[ht]
    \centering
    \includegraphics[width=\linewidth]{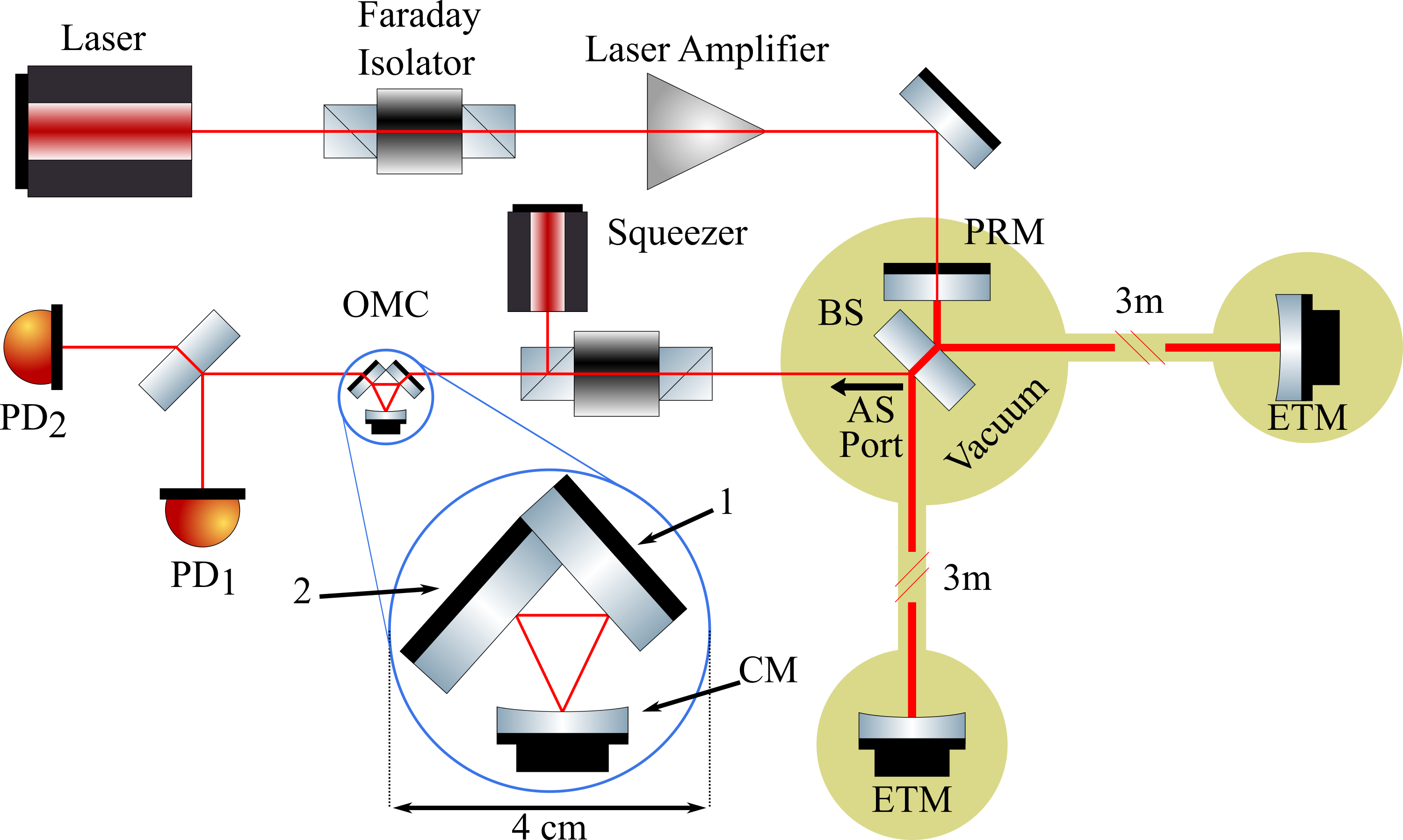}
    \caption{The 0.5 W, 1064 nm laser will be amplified to provide 10 W injection power to the power-recycled Michelson interferometer, which will generate a circulating power of 10 kW. 
    The power recycling mirror (PRM), beam splitter (BS) and end mirrors (ETM) will be in $10^{-8}$ mbar vacuum. At the AS port of the interferometer, the light will be spatially filtered via the OMC and split 50/50 between two high-bandwidth photodetectors. Squeezed vacuum states will be injected into the AS port. The OMC is expanded to emphasise scale. 1 and 2 are the input and output mirrors respectively, and CM is the concave mirror.}
    \label{fig:exp. layout}
\end{figure}

A simplified layout for a single interferometer is displayed in Fig.~\ref{fig:exp. layout}. 

A 1064 nm continuous wave laser with an output power of 0.5 W will be amplified to result in a 10 W injection. We will use a power-recycled Michelson configuration~\cite{schnier_et_al_power_1997} as a strategy to mitigate shot noise, with the expected circulating power reaching $\sim$10 kW. The fundamental TEM$_{\rm 00}$ mode of the carrier will be locked to the common mode arm length of the power-recycled-cavity using a PDH technique~\cite{black_introduction_2001}.  Signals will be obtained using DC-readout~\cite{hild_et_al_dc-readout_2009}, and the output photodetectors (PD$_{1,2}$ of Fig.~\ref{fig:exp. layout})  will be sampled at 500 MHz using high-frequency digitizers. 

\section{Output Mode Cleaner}\label{sec:OMC}

\subsection{Motivation}\label{subsec:OMC motivation}

The DC-readout scheme relies on a small dark-fringe offset, $\delta_{DFO}$, created by an intentional mismatch in arm length such that some carrier leaks out of the anti-symmetric (AS) port to act as a local oscillator. Along with $\delta_{DFO}$, unwanted light in the form of a contrast defect (CD) is also transmitted to the AS port. Contributors to the CD include light due to reflectance and absorption differences between the two arms (causing imperfect destructive interference), and light in the form of higher-order modes~\footnote{TEM$_{\rm 00}$ is the lowest order solution of the paraxial wave equation. An infinite number of higher-order solutions exist which are referred to as HOMs. Compared to TEM$_{\rm 00}$, HOMs pick up additional phase during propagation~\cite{barriga_et_al_optical_2005}.} (HOMs). The CD will be minimised with high quality optics and control systems but some unwanted light is unavoidable. In order to resolve faint signals, $\delta_{DFO}$ must be dominant over the CD, but increasing it introduces new noise sources; in particular, $\delta_{DFO}$ magnitude instability and laser power noise. Excess CD power at the AS port is problematic because it contributes to shot noise, but not signal, and the high-bandwidth PDs risk saturation.

An output mode cleaner (OMC) is a cavity designed to optically filter the laser light transmitted at the output of an interferometer. TEM$_{\rm 00}$ light is resonant while HOMs are not, and are thus suppressed. The inclusion of an OMC $a)$ reduces the required $\delta_{DFO}$; $b)$ facilitates maximising the circulating power in the interferometer; and $c)$ eliminates non-signal-carrying power that otherwise contributes to shot noise. Our upper limit target for CD due to TEM$_{\rm 00}$ is $<10^{-6}$, and for HOMs it's $<10^{-5}$. Fermilab's Holometer's~\cite{chou_et_al_holometer_2017} lowest recorded CD was $\sim 2 \cdot 10^{-5}$.

\subsection{Design fundamentals}\label{subsec:Dsesign}
Here we discuss the process of reaching our design specifications, summarised in Table~\ref{tab:design spec}. The principal function of the OMC is HOM suppression; if $\rm T_{\rm 00}$ is the power transmission of the fundamental mode, the ratio of the power transmission of the ${\rm mn}^{th}$ mode is~\cite{barriga_et_al_optical_2005}
\begin{equation}
    \frac{\rm T_{\rm mn}}{\rm T_{\rm 00}}=\bigg[1+\bigg(\frac{2\mathcal{F}}{\pi}\bigg)^2\sin^2\Big((\text{ m}+\text{n})\text{acos}(\sqrt{g_{\rm cav}})\Big)\bigg]^{-1}.
    \label{eq:triangle hom transmission}
\end{equation}
$\mathcal{F}$ is the cavity finesse and $g_{\rm cav}$ the cavity g-factor. 
\begin{equation}
    \mathcal{F} = \frac{\Delta \nu_{\rm FSR}}{\Delta\nu_{\rm cav}}; \quad \Delta \nu_{\rm FSR} = \frac{c}{\text{L}_{\rm rt}}; \quad g_{\rm cav}=1-\frac{\text{L}_{\rm rt}/2}{\text{ROC}_{\rm CM}},
\label{eq:finesse and fsr and g}
\end{equation}
where $\Delta \nu_{\rm FSR}$ is the cavity free spectral range, $\Delta \nu_{\rm cav}$ the cavity bandwidth, $\text{L}_{\rm rt}$ the cavity's optical round trip length and ROC$_{\rm CM}$ the radius of curvature of the concave mirror (CM). 

\begin{figure}[ht]
    \centering
    \includegraphics[width=\linewidth]{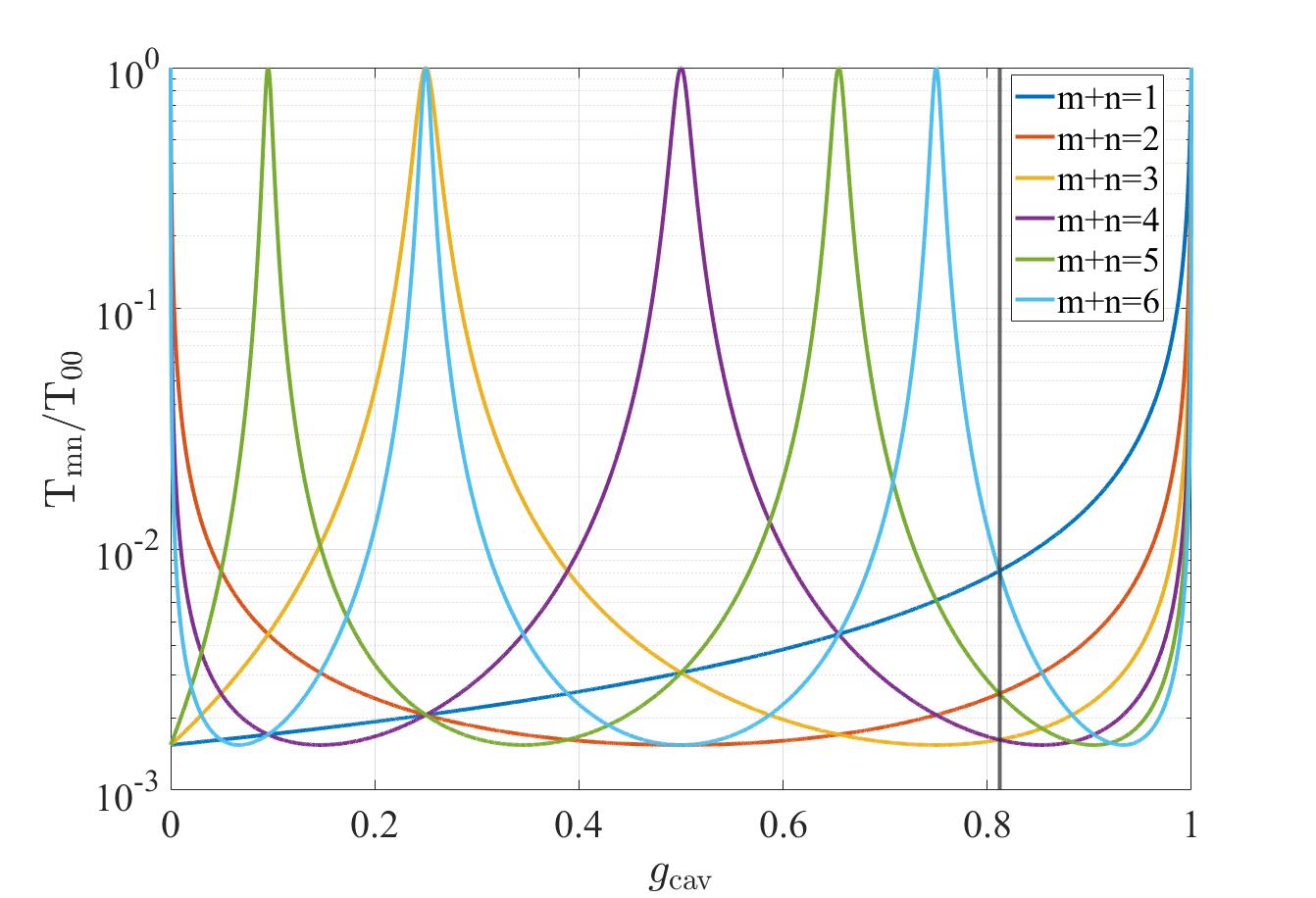}
    \caption{The expected power transmission for the first 6 higher-order modes (Eq. \ref{eq:triangle hom transmission}). Finesse is altered until local minimums reach the desired level, which in our case is $\sim2$ orders of magnitude. This is achieved as displayed, at $\mathcal{F}$ = 40. Appropriate local minimums then reveal the acceptable cavity g-factors. The x line shows the specification g-factor.}
    \label{fig:HOM tx F=40}
\end{figure}
Choosing finesse is a matter of ensuring the local minimums of T$_{\rm mn}$/T$_{\rm 00}$ are below a desired level. Fig.~\ref{fig:HOM tx F=40} shows transmission for the first 6 HOMs, as a function of $g_{\rm cav}$. We sought g-factor values which give $\sim2$ orders of magnitude HOM suppression; there are 4 acceptable minimums in this example, where $\mathcal{F}=40$. 

Cavity bandwidth was chosen to roughly match the useful operating limit of the high-bandwidth PDs, but it is not a strict limit; it can be increased by decreasing the round trip length. It also is not a hard cut-off; signals with a frequency greater than the cavity bandwidth will not be immediately lost.

With finesse and cavity bandwidth, the round trip length is determined. Finally, the ROC$_{\rm CM}$ is calculated upon selecting the g-factor value. The g-factor also governs the waist size and the amount of power coupled to the counter-rotating field within the OMC. In both cases, a higher g-factor is beneficial since a larger waist is easier to maintain, and less power is lost to the counter-rotating field. 

Increasing the angle of incidence (AOI) on the concave mirror couples more power to HOMs since the astigmatism in the waist is worsened (therefore mode matching is deteriorated~\footnote{Models using 2 concave mirrors as an astigmatic telescope have been trialed. So long as AOI$_{\rm CM}$ is kept small (minimising the astigmatism), the power recovered is not worth the added complexity and cost. Though it is an interesting avenue to re-explore if required.}). It also increases the HOM spacing in the sagittal and transverse planes of the cavity, reducing the effective suppression. There is a lower limit to the AOI, due to the  bi-directional reflectance distribution, which generates the counter-rotating field. This is minimised at AOI $>5^{\circ}$.

Cavity finesse also depends on the reflectivities of the mirrors as~\cite{barriga_et_al_optical_2005}
\begin{equation}
    \mathcal{F} \approx \frac{\pi\sqrt{r_1r_2r_{\rm CM}}}{1-r_1r_2r_{\rm CM}},
\label{eq:finesse (reflectivitiy)}
\end{equation}
where $r_{1,2,{\rm CM}}$ are the field amplitude reflectivities of the input, output and concave mirrors respectively. Finesse is already known and, since no light transmits through the concave mirror, it's reflectivity should be as high as practically possible to avoid loss. The input and output mirror reflectivities, which must be equal for an impedance matched cavity, is then determined.

\subsection{Additional considerations}\label{subsec:OMC Aux}

\begin{table}
\caption[]{OMC design specifications}
\label{tab:design spec}
\begin{center}
\begin{tabular}{c|c|c|c|c|c|c|c|c}
$\mathcal{F}$ &$\Delta \nu_{\rm cav}$ & $\Delta \nu_{\rm FSR}$ & $\text{L}_{\rm rt}$ &ROC$_{\text{CM}}$ & Waist &  AOI$_{\rm CM}$ & $r_{1,2}^2$ & $r_{\rm CM}^2$
\\ \hline
40&200 MHz & 8GHz&3.75 cm &100 mm&115 $\mu$m&6$^{\circ}$  & 92.5\% &99.99\%
\\ 
\end{tabular}
\end{center}
\end{table}

For a given physical footprint, the bow-tie geometry produces $\sim$ double the round trip to that of the triangular geometry. With such a small round trip length, a \textit{non-monolithic} bow-tie cavity would be impractical. A linear cavity is also not desirable since reflection is then directed back into the interferometer's AS port.

A monolithic design was considered, where control of round trip length for locking would be achieved via altering the refractive index with temperature. But, a piezoelectronic actuator glued to the concave mirror was decided to be a more straightforward (and well established) method. It will also facilitate a faster response, compared to varying the temperature of a monolithic block.

The flat input and output mirrors will be so close together that individual mounting and alignment control is impractical. Instead, we have decided to have a wedge cut from the edge of the output mirror, such that it can be glued to, or otherwise held in contact with the input mirror (shown in the expanded view, Fig. \ref{fig:exp. layout}) in order to maintain a stable angle.

\section{Projected sensitivity}\label{sec:Projected sensitivity}
\begin{figure}[ht]
    \centering
    \includegraphics[width=\linewidth]{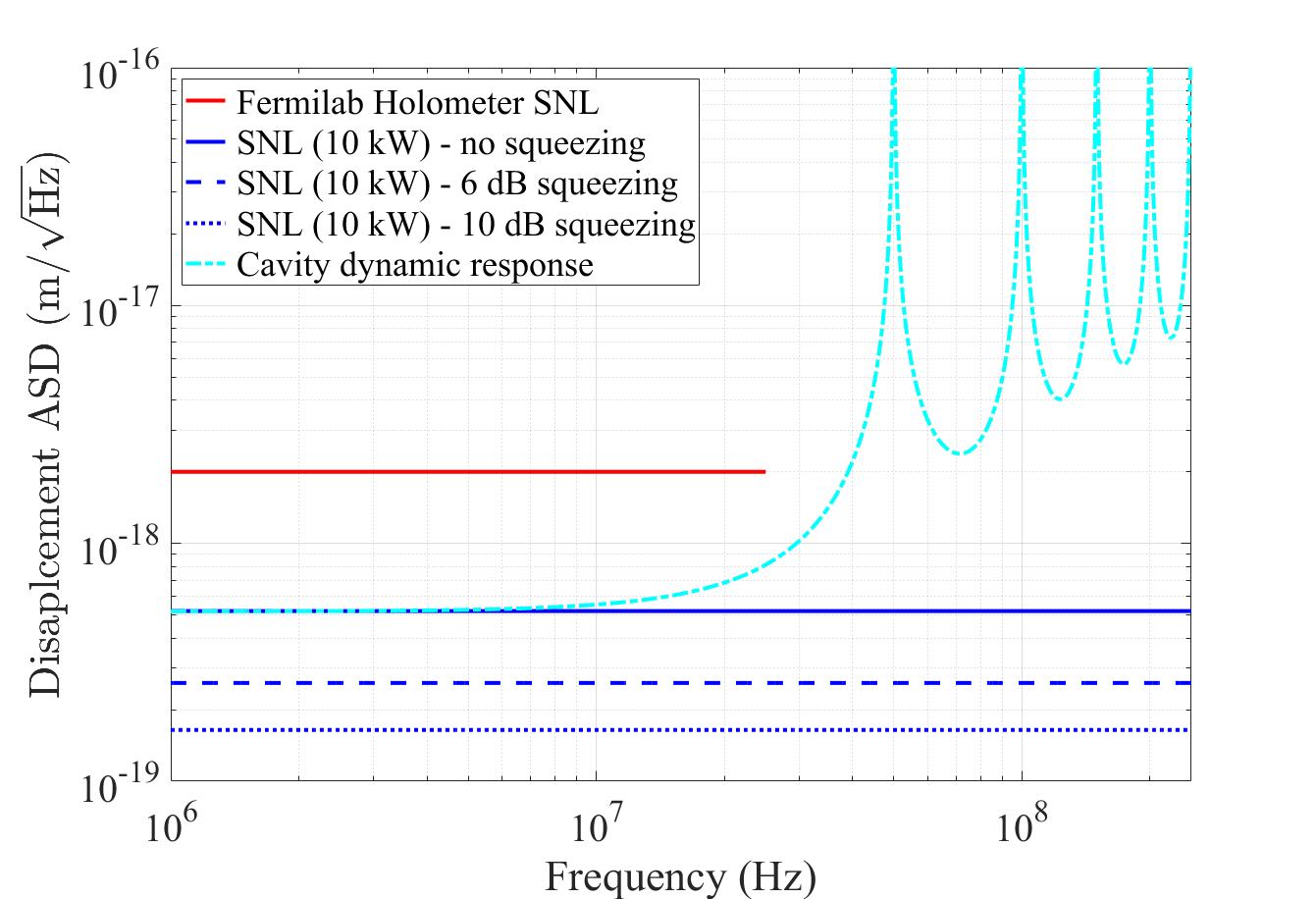}
    \caption{The blue lines show the estimated shot-noise-limited displacement noise ASD of Cardiff's co-located interferometers, each assume 10 kW of power on the beam splitter. Major reasons for Cardiff's increased sensitivity over the Fermilab Holometer (red) include the higher circulating power, the output mode cleaner,  and the injection of squeezed vacuum states. The cyan curve is an equivalent displacement sensitivity, taking into account phenomena that are sensitive to the interferometer frequency response.}
    \label{fig:Projected sensitivity}
\end{figure}

Our co-located interferometers are designed to be shot-noise-limited in the 1--250 MHz band. Other than modal frequency peaks excited by the thermal noise of the optics, all remaining noise sources will be mitigated below this shot noise limit. For a single interferometer without squeezing, the estimated shot-noise-limited displacement ASD is $\sim~5~\times~10^{-19}$~m~/~$\sqrt{\mbox{Hz}}$. This is shown in Fig.~\ref{fig:Projected sensitivity}, along with the ASD for other levels of squeezing.
\section*{Acknowledgments}

The authors are grateful for support from the Science and Technology Facilities Council (STFC), grants ST/T006331/1, ST/I006285/ 1, and ST/L000946/1, the Leverhulme Trust, grant RPG-2019-022 and Cardiff University. 

\section*{References}
\bibliography{references.bib}

\end{document}